\documentclass[prb,twocolumn,superscriptaddress,amsmath,amssymb,floatfix]{revtex4-2}

\usepackage{graphicx}
\usepackage{bm}
\usepackage{dcolumn}
\usepackage[utf8]{inputenc}
\usepackage[hidelinks]{hyperref}

\newcommand{\LLL}{\mathrm{LLL}}
\newcommand{\Dcal}{\mathcal{D}}

\begin{document}

\title{Quantum dynamics of ultrasmall magnetic skyrmions in atomic-defect landscapes}

\author{Samir Lounis}
\email{samir.lounis@physik.uni-halle.de}
\affiliation{Institute of Physics and Halle-Berlin-Regensburg Cluster of Excellence CCE, Martin Luther University Halle-Wittenberg, 06120 Halle (Saale), Germany}

\date{\today}

\begin{abstract}
Magnetic skyrmions are prime candidates for future information technology, and their fate in devices is settled by the defects they encounter. In ultrathin films, skyrmions can be as small as a couple of nanometers, carrying only a few tens of magnetic moments, while atomic defects reshape their energy landscape on the atomic scale. Here, we demonstrate that such ultrasmall skyrmions can behave quantum mechanically and that their quantum dynamics is unusual. Since the gyrotropic force dominates, the two coordinates of a ferromagnetic skyrmion are canonically conjugate and the skyrmion behaves as a particle in the lowest Landau level of an effective magnetic field, with a magnetic length of 0.6~\AA. Utilizing skyrmion-defect interaction profiles of Cr and Fe adatoms and dimers in PdFe/Ir(111), obtained from atomistic spin simulations parametrized from first principles, we unveil quantized gyrotropic orbits with level spacings of 20--85~$\mu$eV, bound not only by pinning defects but also, counter-intuitively, by repulsive ones. Saddle points of the energy landscape act as quantum beam splitters for skyrmions, and quantum depinning reaches rates of 1--100~Hz below the critical current. Embedded impurities, in contrast, bind the skyrmion so strongly that it ceases to behave as a rigid particle, defining a strong-coupling regime. For the Gilbert damping typical of ultrathin films, the quantized levels remain well resolved. The thermal-to-quantum crossover lies between 40 and 160~mK, which is within reach of millikelvin scanning tunneling microscopy, where tip-induced escape statistics and electron spin resonance STM can reveal the quantized gyrotropic motion of skyrmions pinned at defects crafted atom by atom.
\end{abstract}

\maketitle

\section{Introduction}

Magnetic skyrmions~\cite{Bogdanov1989,Roessler2006} are non-collinear spin-textures with particle-like properties, which are  under intense scrutiny for their potential as bits of information in future spintronic devices~\cite{Fert2017,Nagaosa2013}. Their nucleation, motion and stability are, however, strongly affected by material inhomogeneities, which are ineluctable in any device~\cite{Romming2013,Hanneken2016,Mueller2015,LimaFernandes2018,Reichhardt2015,ReichhardtRMP2022}. In this context, PdFe/Ir(111) became a playground of choice, since single skyrmions of a few nanometers can be written and deleted with a scanning tunneling microscope (STM)~\cite{Romming2013}. From first principles, it was  established that single $3d$ and $4d$ atoms, either embedded in the Pd overlayer by substituting a Pd atom (inatoms) or deposited on top of it (adatoms), pin or repel skyrmions, following a universal M-shaped dependence on the $d$-band filling reminiscent of the cohesive energies of transition metals~\cite{LimaFernandes2018} (see also Refs.~\cite{Choi2016,Stosic2017} for MnSi and Co/Pt(111)). Building multi-atomic defects atom by atom enriches further the energy landscape, which can host off-centred minima and even coexisting pinning and repulsive domains~\cite{GedeArjana2020}. Upon application of a current, the skyrmion experiences a gyrotropic motion, clockwise around pinning defects and counter-clockwise around repulsive ones~\cite{LimaFernandes2020JPCM}.

All of these investigations treat the skyrmion as a classical object, and so does the rich body of work on pinning, depinning and creep reviewed recently in Ref.~\cite{ReichhardtRMP2022}. Thermally activated hopping and creep were described utilizing stochastic Thiele equations~\cite{Lin2013b,Troncoso2014} and observed experimentally~\cite{Luo2020,ZhaoL2020}. In contrast, the quantum regime remains largely unexplored. Interestingly, the skyrmions under scrutiny are ultrasmall, with a diameter of about 2~nm and only a few tens of Fe moments, while the skyrmion-defect interaction profiles vary on the scale of a few \AA ngstr\"om. A natural question arises then: is the motion of such a skyrmion still classical, or does its position acquire a quantum-mechanical nature? Quantum aspects of skyrmions were mostly addressed with model potentials. For example, the quantized motion of a skyrmion in a pinning well~\cite{Lin2013,Psaroudaki2017}, its quantum depinning out of an impurity potential~\cite{PsaroudakiLoss2020}, the collapse of skyrmions~\cite{DerrasChouk2018,Vlasov2020}, quantum skyrmion liquids and frustrated quantum magnets~\cite{Takashima2016,Lohani2019}, as well as skyrmion qubits based on the helicity degree of freedom~\cite{Psaroudaki2021,PsaroudakiPanagopoulos2022,PsaroudakiAPL2023}. However, the connection to realistic defects with a well-defined chemical nature, which can be designed and manipulated atom by atom, is terra incognita.

\begin{figure*}[t]
\includegraphics[width=\textwidth]{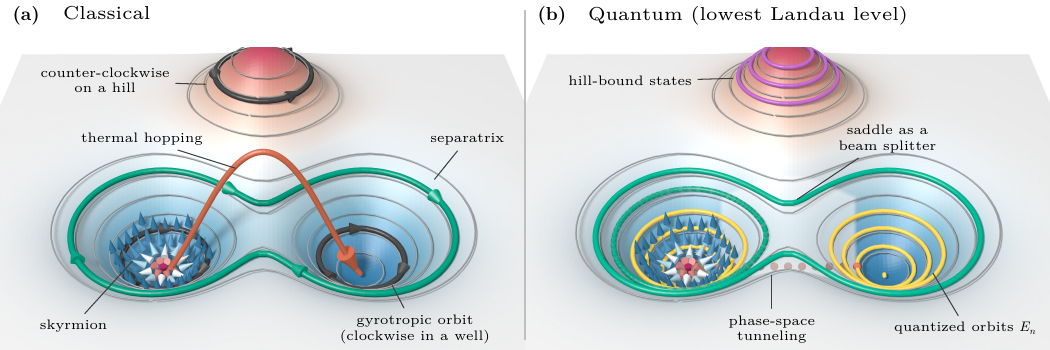}
\caption{Schematic comparison of classical and quantum skyrmion dynamics in an energy landscape crafted by atomic defects, here two pinning wells (blue) connected by a saddle point and a repulsive hill (red). (a) Classically, the gyrotropic force drives the skyrmion along equipotential lines: it orbits clockwise in a well and counter-clockwise around a hill, a trajectory passing close to the saddle follows the separatrix deterministically, and transitions between wells occur by thermal hopping over the barrier. (b) In the lowest Landau level, the orbits are quantized into discrete states $E_n$, located in the wells but also on top of the repulsive hill. At the saddle point, an incoming skyrmion is split into two outgoing paths, similarly to an electron at a quantum point contact, and the skyrmion escapes from a pinning centre by tunneling in phase space.}
\label{fig:concept}
\end{figure*}

The key to the answer lies in the gyrotropic (Magnus) force entering the Thiele equation~\cite{Thiele1973}. In contrast to a massive particle, the dynamics of a ferromagnetic skyrmion is dominated by this force, which acts as a Lorentz force on a charged particle in a strong magnetic field. Quantum mechanically, the skyrmion is thus a particle living in the lowest Landau level (LLL). It has no kinetic energy, and the energy landscape generated by the defect becomes its Hamiltonian. This has three remarkable consequences, which define the backbone of this work: (i)~the skyrmion can be bound by repulsive defects as well as by pinning ones; (ii)~at saddle points of the landscape, an incoming skyrmion is split into two outgoing paths, similarly to an electron at a quantum point contact; (iii)~close to the depinning threshold, the skyrmion escapes from a pinning centre by tunneling in phase space rather than through a barrier. These three regimes are illustrated schematically in Fig.~\ref{fig:concept}.

Here, we develop this picture and combine it with the skyrmion-defect interaction profiles of PdFe/Ir(111) obtained from reported first-principles-based investigations~\cite{LimaFernandes2018,LimaFernandes2020JPCM,GedeArjana2020}. After introducing the theory in Sec.~\ref{sec:theory} and the energy landscapes in Sec.~\ref{sec:landscapes}, we address the quantized gyrotropic orbits in Sec.~\ref{sec:orbital}, the saddle points in Sec.~\ref{sec:saddle} and the quantum depinning in Sec.~\ref{sec:depinning}. Possible experimental signatures are discussed in Sec.~\ref{sec:experiment} before concluding. We note that Lin and Bulaevskii~\cite{Lin2013} quantized the skyrmion motion in a model pinning well, while Psaroudaki and Loss~\cite{PsaroudakiLoss2020} predicted the quantum depinning of a skyrmion out of a model impurity potential, with a crossover temperature of about 20~mK. In contrast, we consider realistic wells, hills, ring-like wells and saddle points, anchor the energy scale to atomistic dynamics and establish the link to defects already explored theoretically and experimentally.

\section{Skyrmions as lowest-Landau-level particles}
\label{sec:theory}

We describe the skyrmion by the position $\mathbf R=(X,Y)$ of its centre, while its internal deformations are assumed to be fast. The dynamics is then dictated by the collective-coordinate Lagrangian
\begin{equation}
L = \frac{G}{2}\,(X\dot Y - Y\dot X) + \frac{M}{2}\,\dot{\mathbf R}^2 - U(\mathbf R),
\qquad G = 4\pi Q\,\frac{\hbar S}{A_{\mathrm{cell}}},
\label{eq:L}
\end{equation}
where $Q=-1$ is the topological charge, $S$ the spin per magnetic unit cell of area $A_{\mathrm{cell}}$, $M$ an inertial mass induced by the deformation modes~\cite{Makhfudz2012,Buettner2015,Schutte2014} and $U(\mathbf R)$ the energy landscape generated by the defect. After adding the Gilbert damping $\alpha$ and a spin-transfer torque with drift velocity $\mathbf u$, one recovers the Thiele equation utilized in our atomistic simulations~\cite{LimaFernandes2020JPCM,Thiele1973}:
\begin{equation}
\mathbf G \times (\mathbf u - \dot{\mathbf R}) + \Dcal\,(\beta\mathbf u - \alpha\dot{\mathbf R}) - M\ddot{\mathbf R} = \nabla U(\mathbf R),
\label{eq:Thiele}
\end{equation}
with $\Dcal$ the dissipation constant and $\beta$ the non-adiabatic parameter.

The first term of Eq.~\eqref{eq:L} is at the heart of the peculiar behavior of skyrmions. It has the same form as the coupling of a charged particle to a vector potential and dominates over the inertial term for ferromagnetic skyrmions. Neglecting the inertia, $X$ and $Y$ become canonically conjugate, and the quantization leads to
\begin{equation}
[\hat X, \hat Y] = i\,\ell^2\,\mathrm{sgn}(Q), \qquad \ell^2 = \frac{\hbar}{|G|} = \frac{A_{\mathrm{cell}}}{4\pi S}.
\label{eq:comm}
\end{equation}
Interestingly, the magnetic length $\ell$ is solely determined by the spin density. For PdFe/Ir(111), with $a=2.715$~\AA, $A_{\mathrm{cell}}=6.38$~\AA$^2$ and $S=1.35$ as obtained from the Fe moment of $2.7\,\mu_B$, we find $\ell=0.61$~\AA. Thus, a quantum state of the skyrmion position occupies an area $2\pi\ell^2\approx2.4$~\AA$^2$, i.e.\ a fraction of an atomic cell. While small, this area is not negligible in comparison to energy landscapes varying over a few lattice constants, which is the reason why quantum effects can be at play.

In the absence of inertia, the Hamiltonian reduces to the energy landscape with operator arguments:
\begin{equation}
\hat H_{\LLL} = U(\hat X, \hat Y),
\label{eq:HLLL}
\end{equation}
which we consider in Weyl ordering, such that the computed $U(\mathbf R)$ is its classical symbol. Classically, Eq.~\eqref{eq:HLLL} drives the skyrmion along equipotential lines, never downhill. Around an extremum characterized by the curvatures $k_x$ and $k_y$, the skyrmion follows an elliptical orbit, clockwise in a well and counter-clockwise on a hill for $Q=-1$, as found in our atomistic spin-dynamics simulations~\cite{LimaFernandes2020JPCM}. Quantum mechanically, the extremum hosts a ladder of equidistant states:
\begin{equation}
E_n = U_0 \pm \hbar\omega\left(n + \tfrac12\right), \qquad \hbar\omega = \ell^2\sqrt{k_x k_y},
\label{eq:ladder}
\end{equation}
with the upper (lower) sign for wells (hills). The level spacing is given by the curvature of the energy landscape multiplied by $\ell^2$ and defines a crossover temperature $T^*=\hbar\omega/(2\pi k_B)$, below which the quantum fluctuations of the skyrmion position overcome the thermal ones.

Two effects can, however, alter this picture. First, the LLL description holds only if $\hbar\omega$ is smaller than the energies of the internal modes of the skyrmion, which play the role of the cyclotron gap. For skyrmions of about 2~nm in PdFe/Ir(111), our atomistic spin-dynamics calculations place the lowest internal mode at about 0.4~meV. These internal modes are magnon states bound to the skyrmion, which lie below the gap of the ferromagnetic background, $(2K+\mu B)/S$, set by the magnetic anisotropy $K$ and the Zeeman energy; with the larger anisotropy obtained from first principles~\cite{LimaFernandes2018}.

We note that the skyrmion-defect landscapes $U(\mathbf R)$ are obtained after relaxing the spin texture for each skyrmion position, such that the internal deformations are eliminated adiabatically. This is justified as long as the dimensionless coupling
\begin{equation}
\eta=\frac{\hbar\omega}{\hbar\omega_{\mathrm{int}}}=\frac{\ell^2\sqrt{k_xk_y}}{\hbar\omega_{\mathrm{int}}}
\label{eq:eta}
\end{equation}
remains small, with $\hbar\omega_{\mathrm{int}}$ the energy of the lowest internal mode. Close to the depinning threshold, the curvature of the tilted well decreases, such that the rigid description improves rather than deteriorates. Second, the Gilbert damping acts as an Ohmic bath, forcing the orbits to spiral inward in wells and outward on hills with a rate
\begin{equation}
\Gamma = \frac{2\alpha\Dcal\,|G|}{G^2 + \alpha^2\Dcal^2}\,\omega, \qquad Q_{\mathrm f} = \frac{\omega}{\Gamma} \approx \frac{|G|}{2\alpha\Dcal}.
\label{eq:Q}
\end{equation}
The quantized levels are resolved as long as the quality factor $Q_{\mathrm f}$ is larger than one. For ultrathin metallic films hosting skyrmions, the Gilbert damping typically lies in the range $\alpha\approx0.01$--0.1, 
and $\alpha=0.05$ was used in our spin-dynamics simulations of PdFe/Ir(111)~\cite{LimaFernandes2020JPCM}. Since $\Dcal/|G|\approx1$ for compact skyrmions~\cite{LimaFernandes2020JPCM}, one obtains $Q_{\mathrm f}\approx5$--50, such that the quantized levels are well resolved. In tunneling processes, the same bath adds a term of the order of $(\alpha\Dcal/|G|)(\Delta/\ell)^2$ to the exponent, with $\Delta$ the tunneling distance~\cite{Caldeira1983}. For $\alpha\Dcal/|G|\lesssim0.1$ and tunneling distances of 1--2~\AA, this correction amounts to at most about one, i.e.\ to a few percent of the exponents discussed in Sec.~\ref{sec:depinning}.

\section{Skyrmion-defect interaction profiles}
\label{sec:landscapes}

The only input needed by the theory is the energy landscape $U(\mathbf R)$. Previous investigations of PdFe/Ir(111) offer two complementary sources. From first principles, utilizing the full-potential relativistic Korringa--Kohn--Rostoker (KKR) Green function method, it was  established how single $3d$ and $4d$ atoms embedded in the Pd overlayer interact with a skyrmion~\cite{LimaFernandes2018,Papanikolaou2002,JuKKR2022}. There, a skyrmion of 2.2~nm made of 37 Fe atoms is created without magnetic field by an embedding technique: the moment at its centre is flipped with respect to the ferromagnetic background, and the non-collinear texture is relaxed self-consistently in the presence of the impurity. Interestingly, these calculations already reveal all landscape topologies distinguished by the theory: Tc and Ru induce centred wells, V, Cr and Co sharp centred hills, while Mn, Mo and Nb give rise to ring-like wells surrounding the impurity. The binding energies, given per Fe atom of the skyrmion, amount in total to tens and up to 170~meV, i.e.\ to a sizeable fraction of the skyrmion energy itself, about 410~meV, and vary on the scale of a lattice constant. As shown in Sec.~\ref{sec:orbital}, such strongly interacting embedded impurities drive the skyrmion out of the regime of a rigid particle.

We therefore base our quantitative analysis on atomistic spin simulations, parametrized with the \textit{ab initio} tensor of magnetic interactions, which provide two-dimensional skyrmion-defect interaction maps for single Cr and Fe adatoms and for their nearest-neighbor dimers at 10~T~\cite{GedeArjana2020}. The energies are given per impurity; for a single adatom, this is the total energy difference between the skyrmion located at the defect and far away from it, while for dimers the energies are multiplied by the number of adatoms. These landscapes are remarkably rich. The Cr and Fe adatoms generate repulsive hills of 16.8 and 4.8~meV, respectively, while the dimers host two minima, 9.7 and 2.1~meV deep, connected by a saddle point, complemented by a hill in the case of Fe. The same model was utilized in our spin-dynamics simulations of current-driven skyrmions~\cite{LimaFernandes2020JPCM}, in which the energies are not renormalized, such that the classical limit of our theory can be directly confronted with atomistic dynamics. The published maps were interpolated from calculations on a grid of skyrmion positions. 
The profiles utilized in this work are shown in Figs.~\ref{fig:inatoms}(a) and \ref{fig:inatoms}(b) for single defects and in Fig.~\ref{fig:saddles} for the dimers.

\section{Quantized gyrotropic orbits: pinning wells versus repulsive hills}
\label{sec:orbital}

We start with the simplest landscapes, i.e.\ single defects hosting a single extremum. For a rotationally symmetric $U(r)$, the problem can be solved exactly. The eigenstates are angular-momentum orbitals $|m\rangle$ circulating around the defect at a radius $r_m\approx\ell\sqrt{2m+1}$, and their energies are obtained from their Wigner functions (see Appendix~\ref{app:wigner}):
\begin{equation}
\begin{split}
E_m &= \int d^2r\, W_m(r)\, U(r),\\
W_m(r) &= \frac{(-1)^m}{\pi\ell^2}\, e^{-r^2/\ell^2}\, L_m\!\left(\frac{2r^2}{\ell^2}\right).
\end{split}
\label{eq:Wigner}
\end{equation}
In other words, the spectrum samples the energy landscape on rings, each with an area of $2\pi\ell^2$. Around an extremum, Eq.~\eqref{eq:Wigner} reduces to the ladder of Eq.~\eqref{eq:ladder}, with curvatures obtained from the two-dimensional maps. The resulting level spacings are collected in Table~\ref{tab:maps}.

\begin{table*}[t]
\caption{Level spacings $\hbar\omega$, frequencies and crossover temperatures $T^*$ obtained from the two-dimensional interaction maps of Ref.~\cite{GedeArjana2020} (total energies, fcc stacking). Central values and ranges reflect the uncertainty due to the interpolation of the published maps.}
\label{tab:maps}
\begin{ruledtabular}
\begin{tabular}{llccc}
Defect & Extremum & $\hbar\omega$ ($\mu$eV) & Frequency (GHz) & $T^*$ (mK) \\
\hline
Fe adatom & hill, $+4.8$~meV & 24 (20--29) & 5.9 & 45 (36--53) \\
Cr adatom & hill, $+16.8$~meV & $\lesssim 40$\footnotemark[1] & $\lesssim 10$ & $\lesssim 75$ \\
Fe dimer & wells, $-2.1$~meV & 40 (39--44) & 9.7 & 74 (72--81) \\
Cr dimer & wells, $-9.7$~meV & 84 (73--88) & 20 & 154 (135--162) \\
\end{tabular}
\end{ruledtabular}
\footnotetext[1]{The top of the Cr hill is flattened by the saturation of the published color scale, which only allows an upper bound.}
\end{table*}

The pinning wells of the dimers bind the skyrmion with level spacings of 40 and 84~$\mu$eV, corresponding to crossover temperatures of 75 and 155~mK, while the repulsive adatoms host hill-bound states with spacings of 20--40~$\mu$eV, as discussed below. Embedded impurities behave differently. Being located within the Pd overlayer, they hybridize directly with the Fe atoms hosting the skyrmion and interact much more strongly with it than adatoms do~\cite{LimaFernandes2018}. This places them in a strong-coupling regime, beyond the picture of a rigid skyrmion: while $\eta\approx0.05$--0.2 for the adatoms and dimers of Table~\ref{tab:maps}, $\eta$ reaches 5--100 for the embedded impurities. According to ab-initio simulations, the binding energies of embedded impurities reach $-171$ and $-98$~meV for Tc and Ru and 126--141~meV for V, Cr and Co, i.e.\ a sizeable fraction of the skyrmion energy, while varying on the scale of a lattice constant. The corresponding gyrotropic level spacings would amount to 2--4~meV for the wells and up to tens of meV for the hills, exceeding the internal modes by one to two orders of magnitude. Orbital motion and internal deformations are then intimately coupled, in line with the strongly asymmetric skyrmion found at the vicinity of these impurities~\cite{LimaFernandes2018}, and the skyrmion can no longer be treated as a rigid particle. A simple estimate, including the inertia associated with the internal modes, places the quantized levels at the geometric mean of the gyrotropic and internal-mode energies, i.e.\ at about 1~meV for Tc and Ru, which would correspond to crossover temperatures of several kelvin. We note that the embedding technique fixes the region hosting the skyrmion, such that this estimate is only indicative. A quantitative description of this strong-coupling regime requires a joint quantization of the position and of the internal modes of the skyrmion, which we leave for future work.

\begin{figure*}[t]
\includegraphics[width=\textwidth]{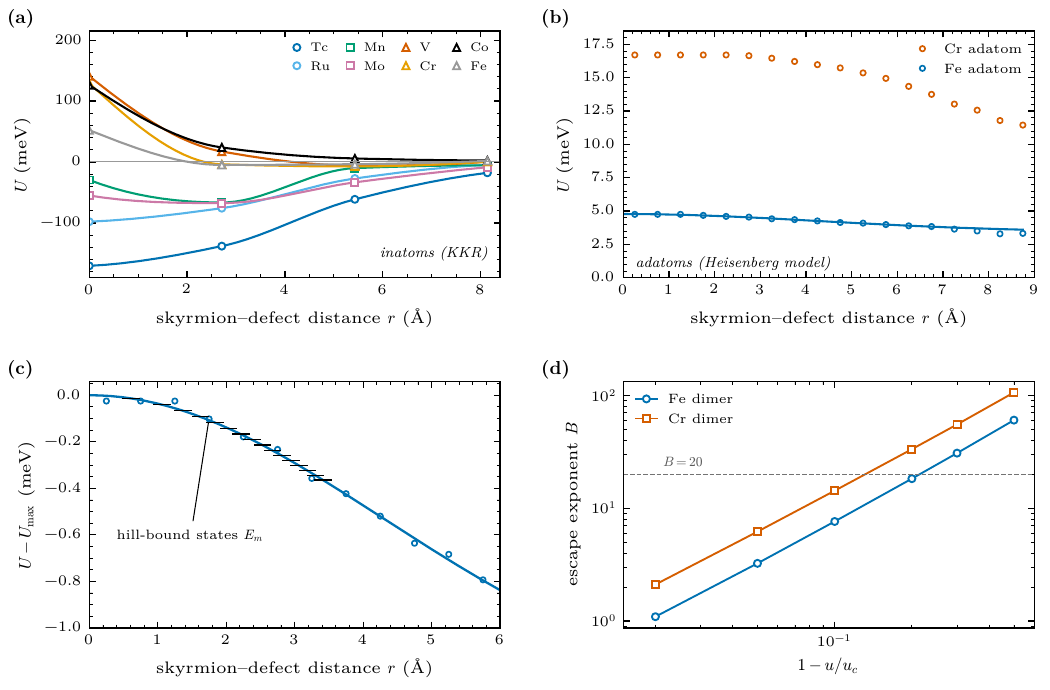}
\caption{(a) Skyrmion--inatom interaction energies obtained from first-principles~\cite{LimaFernandes2018}, i.e. KKR method, (symbols) as a function of the distance between the skyrmion core and the impurity, given as total energies, with shape-preserving interpolations (lines). (b) Skyrmion--adatom interaction energies obtained with the \textit{ab initio}-parametrized  Heisenberg model~\cite{GedeArjana2020}, radially averaged over the directions covered by the published maps. (c) Interaction energy of the Fe adatom close to the top of the hill (symbols) and its smooth fit (line), together with the lowest hill-bound orbital levels $E_m$ (bars), placed at the orbital radius $r_m=\ell\sqrt{2m+1}$; the level spacing amounts to about 28~$\mu$eV. (d) Quantum-depinning exponent $B$ for the escape of a skyrmion out of one well of the Fe and Cr dimers over their saddle point, as a function of the distance $1-u/u_c$ from the classical threshold. The dashed line marks $B=20$, corresponding to escape rates of a few tens of Hz. Panel (a) is adapted from Ref.~\cite{LimaFernandes2018} and panels (b) and (c) from Ref.~\cite{GedeArjana2020}, both published under a Creative Commons Attribution 4.0 International License (http://creativecommons.org/licenses/by/4.0/); the data were converted to total energies, radially averaged and re-plotted}
\label{fig:inatoms}
\end{figure*}

The case of repulsive defects is particularly intriguing, since the LLL physics contradicts here the classical intuition. On a hill, the ladder of Eq.~\eqref{eq:ladder} is inverted, $E_n=U_{\max}-\hbar\omega(n+\frac12)$, with states circulating around the top of the hill, as illustrated for the Fe adatom in Fig.~\ref{fig:inatoms}(c). In the absence of damping, these states are as stationary as those hosted by a well, since no kinetic energy allows the skyrmion to roll down the hill. Such states do not exist for a massive particle. The damping, however, lowers the energy, which leads to a relaxation toward the bottom of a well but to an outward spiraling motion on a hill. Therefore, a skyrmion positioned on top of a repulsive defect is metastable, with a lifetime dictated by the damping rather than by the shape of the hill. With level spacings of 20--40~$\mu$eV, the Fe and Cr adatoms host such hill-bound states with crossover temperatures of 40--75~mK.

Finally, we note that ring-like wells, as obtained in the KKR calculations for Mn, Mo and Nb inatoms~\cite{LimaFernandes2018}, would host their lowest states on a ring around the impurity, in orbitals with large angular momentum. Since neighboring orbitals are radially displaced by only $\ell^2/r$, they are almost degenerate, and the threefold symmetry of the lattice site, coupling $m$ to $m\pm3$, should define the actual level structure. Interestingly, off-centred pinning is precisely what is observed experimentally for inlayer defects in PdFe/Ir(111), where each defect locks the skyrmion at a fixed off-centre position~\cite{Hanneken2016}.

\section{Saddle points as quantum beam splitters for skyrmions}
\label{sec:saddle}

Once a landscape hosts more than one extremum, saddle points naturally emerge, and it is precisely there that the quantum nature of a LLL particle is most pronounced. Classically, a skyrmion drifting along an equipotential line reaches the saddle and continues along one of two branches, depending on the side of the separatrix it comes from. Quantum mechanically, this choice becomes probabilistic. Close to the saddle, $U\approx U_s+(k_\parallel x^2-k_\perp y^2)/2$, and a rescaling preserving Eq.~\eqref{eq:comm} maps Eq.~\eqref{eq:HLLL} onto an inverted harmonic oscillator. The latter is exactly solvable~\cite{Fertig1987}, and the probability that the skyrmion leaves along the classically forbidden branch reads
\begin{equation}
T(E) = \left[1 + e^{-2\pi(E - U_s)/\varepsilon_s}\right]^{-1}, \quad \varepsilon_s = \ell^2\sqrt{k_\parallel k_\perp}.
\label{eq:TE}
\end{equation}
Thus, the classical step at $E=U_s$ is smeared over $\varepsilon_s/2\pi$, and at $E=U_s$ the skyrmion is equally split between the two branches. This is the skyrmionic analogue of a quantum point contact. The quantum smearing overcomes the thermal one below $T_s=\varepsilon_s/(2\pi k_B)$.

\begin{figure*}[t]
\includegraphics[width=\textwidth]{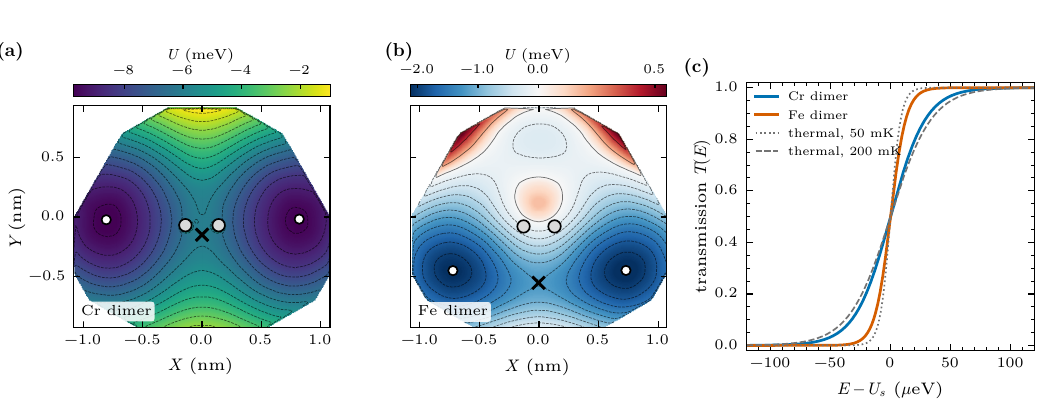}
\caption{Skyrmion--defect interaction landscapes of the nearest-neighbor (a) Cr and (b) Fe dimers deposited on the fcc sites of PdFe/Ir(111), obtained from the atomistic spin simulations of Ref.~\cite{GedeArjana2020} (total energies). The adatoms are shown as grey circles, the minima as white circles and the saddle points as crosses. (c) Probability $T(E)$ for a skyrmion to leave a saddle along the classically forbidden branch, Eq.~\eqref{eq:TE}, for the Cr ($\varepsilon_s=93~\mu$eV, upper end of its range) and Fe ($\varepsilon_s=41~\mu$eV) dimers, compared with thermal steps of width $k_BT$ at 50 and 200~mK. Panels (a) and (b) are adapted from Ref.~\cite{GedeArjana2020}, published under a Creative
Commons Attribution 4.0 International License (http://creativecommons.org/licenses/by/4.0/); the maps were converted
to total energies and re-plotted, and the adatoms, minima and saddle points were redrawn.}
\label{fig:saddles}
\end{figure*}

The nearest-neighbor Cr and Fe dimers offer two ideal examples (see Fig.~\ref{fig:saddles}). Each of them hosts two minima separated by about 1.5~nm and connected by a saddle point. We note that direct tunneling between the two minima is hopeless, since it is suppressed as $\exp(-A/\ell^2)$, with $A$ the area separating the two orbits; every \AA$^2$ costs a factor $e^{-2.7}$. Thus, the skyrmion has to pass through the saddle, where Eq.~\eqref{eq:TE} applies. The saddle parameters are listed in Table~\ref{tab:saddles}. While the Fe saddle is well defined, the Cr saddle is located at the vicinity of the adatoms, where the published map is partly masked by the atom symbols, such that only a range can be given for its curvatures.

\begin{table}[t]
\caption{Saddle parameters from the maps of Ref.~\cite{GedeArjana2020} (total energies; $\Delta U$ is the barrier from the minima to the saddle). Ranges reflect the interpolation of the published maps.}
\label{tab:saddles}
\begin{ruledtabular}
\begin{tabular}{lccc}
Dimer & $\Delta U$ & $\varepsilon_s$ & $T_s$ \\
(fcc) & (meV) & ($\mu$eV) & (mK) \\
\hline
Fe dimer & 0.92 & 41 (40--41) & 75 \\
Cr dimer & 3.7 & 60--95 & 110--175 \\
\end{tabular}
\end{ruledtabular}
\end{table}

How can such an effect be detected? In devices allowing lateral currents, the skyrmion drifts upon application of a spin-polarized current along the equipotential lines of the tilted landscape $U(\mathbf R)-\mathbf F\cdot\mathbf R$, with $\mathbf F=G\,\hat{\mathbf z}\times\mathbf u$. The energy of an incoming trajectory is then defined by its impact parameter $b$ with respect to the defect, $E-U_s\approx F(b-b_s)$, where $b_s$ marks the classical separatrix between capture and deflection. This is exactly the scenario explored in previous spin-dynamics simulations~\cite{LimaFernandes2020JPCM}, where a shift of the incoming path by a fraction of a nanometer decides whether the skyrmion is pinned or deflected, as well as in particle-based models, where the Magnus force renders the deflection asymmetric with respect to the impact parameter~\cite{Reichhardt2015b,ReichhardtRMP2022}. Interestingly, Eq.~\eqref{eq:TE} transforms this sharp boundary into a smooth one, with a width
\begin{equation}
\delta b_{\mathrm q} = \frac{\varepsilon_s}{2\pi F} = \frac{\ell^2\varepsilon_s}{2\pi\hbar u}.
\label{eq:db}
\end{equation}
Thermal fluctuations broaden the boundary by $\delta b_{\mathrm{th}}=k_BT/F$, such that the ratio between the two widths is $T_s/T$. For the Fe saddle and a drift velocity of 1~m/s, $\delta b_{\mathrm q}=0.37$~\AA, which increases to 3.7~\AA\ at 0.1~m/s. The signature of the quantum regime is thus a capture window that stops sharpening upon cooling below $T_s$.

Multi-atomic defects host several saddle points, and a skyrmion transmitted at one saddle and reflected at another one encloses an area $A$ and acquires a phase $A/\ell^2$, i.e.\ a full turn per 2.4~\AA$^2$. Two saddles in series could then act as an interferometer, similarly to quantum-Hall Fabry--P\'erot devices. Such an interference requires, however, phase coherence between paths separated by a distance $\Delta$, which is limited by the same Ohmic bath responsible for the damping. In the high-temperature limit of the Caldeira--Leggett model~\cite{Caldeira1983}, the decoherence rate reads $\Gamma_\varphi\approx2(\alpha\Dcal/|G|)(\Delta/\ell)^2k_BT/\hbar$. For saddles separated by about 1.5~nm, $\alpha\Dcal/|G|\approx0.01$--0.1 and $T=50$~mK, this yields coherence times in the picosecond range, much shorter than the time needed to drift between the saddles. Skyrmion interferometry is therefore out of reach in PdFe/Ir(111). We emphasize that the beam splitting at a single saddle, Eq.~\eqref{eq:TE}, does not require coherence between distant paths and is not affected by this limitation.

\section{Quantum depinning of skyrmions}
\label{sec:depinning}

The most accessible quantum effect is probably the escape of the skyrmion from a pinning centre. A current, or within STM a tip-induced tilt of the landscape (Sec.~\ref{sec:experiment}), pushes the skyrmion against the restoring force of the defect, and just below the critical drive the barrier holding the skyrmion becomes small. This is the regime in which macroscopic quantum tunneling was first observed in Josephson junctions~\cite{Martinis1987}, and the same strategy applies here, with a twist. Since the skyrmion lives in the LLL, it tunnels in phase space: the coordinate along the drive plays the role of the position, while the transverse coordinate plays the role of the momentum. Interestingly, such a Hall-type tunneling is reminiscent of the quantum creep of vortices in high-temperature superconductors~\cite{Blatter1994}. The quantum depinning of a skyrmion was predicted for a model impurity potential~\cite{PsaroudakiLoss2020}; here, we address realistic, element-specific pinning centres.

Interestingly, this twist leads to a simple result. If the landscape is harmonic along the transverse direction with a curvature $k_y$, the LLL problem maps exactly onto that of a one-dimensional particle with a mass $1/(k_y\ell^4)$ (see Appendix~\ref{app:escape}). Close to the threshold, the tilted landscape takes the generic form $\lambda(\delta x-x^3/3)$, with $\delta\propto1-u/u_c$ and a barrier $\Delta U_b=\frac43\lambda\delta^{3/2}$, and the escape rate is given by
\begin{equation}
\Gamma_{\mathrm q} \simeq A\,\frac{\omega_w}{2\pi}\, e^{-B}, \qquad B = \frac{36}{5}\,\frac{\Delta U_b}{\hbar\omega_w} \propto \left(1 - \frac{u}{u_c}\right)^{5/4}.
\label{eq:B}
\end{equation}
This is the well-known result for a massive particle in a cubic potential, where the plasma frequency is replaced by the gyrotropic frequency $\hbar\omega_w=\ell^2\sqrt{k_xk_y}$ at the bottom of the tilted well. Since $B$ measures an area in phase space in units of $\ell^2$, it is independent of the normalization of the energies, which removes the main uncertainty discussed in Sec.~\ref{sec:landscapes}. Moreover, the crossover temperature $T^*=\hbar\omega_w/(2\pi k_B)$ depends only weakly on the drive.

We evaluated the escape exponent for the dimers, for which the skyrmion escapes from one well over the saddle point into the second well. Since second derivatives of interpolated maps are noisy, we performed the analysis on smooth fits of the landscapes, given by sums of Gaussians that reproduce the maps within 1--3\%. The drive was applied along the dimer axis and the escape exponent was evaluated along the branch of the metastable minimum up to the threshold [Table~\ref{tab:depinning} and Fig.~\ref{fig:inatoms}(d)].

\begin{table}[t]
\caption{Quantum depinning out of one well of the dimers over their saddle point. $\delta_{20}$ is the value of $1-u/u_c$ at which $B=20$; $\hbar\omega_w$, the rate $\Gamma_{20}$ (without the prefactor $A$) and $T^*$ refer to that drive.}
\label{tab:depinning}
\begin{ruledtabular}
\begin{tabular}{lccccc}
 & $u_c$ & & $\hbar\omega_w$ & $\Gamma_{20}$ & $T^*$ \\
Dimer & (m/s) & $\delta_{20}$ & ($\mu$eV) & (Hz) & (mK) \\
\hline
Fe dimer & 12 & 0.21 & 32 & 15 & 60 \\
Cr dimer & 43 & 0.12 & 62 & 30 & 115 \\
\end{tabular}
\end{ruledtabular}
\end{table}

One notices that both defects behave similarly. When the drive approaches the threshold, $B$ drops from about 50--100 to a few units, and the escape rate increases from far below a millihertz to the megahertz range. Rates of 1--100~Hz, convenient for experiments, are reached 10--25\% below the threshold. Interestingly, the critical drift velocities of 12 and 43~m/s are of the same order as the depinning velocity of about 18~m/s found for the Ru inatom in spin-dynamics simulations~\cite{LimaFernandes2020JPCM}, which illustrates the consistency between the landscapes and the atomistic dynamics.

\section{Experimental signatures}
\label{sec:experiment}

PdFe/Ir(111) is an ideal platform. Skyrmions can be written, deleted, pinned and moved with STM tips~\cite{Romming2013,Hanneken2016}, and the experiments of Ref.~\cite{Hanneken2016} already demonstrate the main ingredients required here. Inlayer defects pin skyrmions at well-defined off-centre positions, single Co adatoms are almost inert, in line with first-principles predictions~\cite{LimaFernandes2018}, while Co clusters pin skyrmions strongly enough to drag them across the surface upon lateral manipulation of the cluster. These experiments were, however, performed at 4.2--8~K, an order of magnitude above the crossover temperatures of 40--160~mK found here, which explains why only classical pinning has been observed so far. Dilution-refrigerator STM is therefore required. We note that the quantum depinning of a related magnetic soliton, the vortex core in permalloy disks, has already been observed experimentally~\cite{Zarzuela2012}. Since the electronic temperature at the junction often exceeds the base temperature, the most promising candidates are the defects with the largest level spacings, i.e.\ the Cr dimer, with $T^*\approx155$~mK, and the Fe dimer with its well-defined saddle point, with $T^*\approx75$~mK.

It is important to relate our estimates to the experimental conditions. Isolated skyrmions in PdFe/Ir(111) are stabilized at 1--3.25~T, with diameters of about 1.8~nm at 3~T and 3.6~nm at 1~T~\cite{Hanneken2016}. The Heisenberg-model landscapes were obtained at 10~T for a skyrmion of about 3~nm~\cite{GedeArjana2020}. It is the size of the skyrmion, rather than the field itself, that determines the curvature of the landscapes, and this size corresponds to experimental skyrmions at about 1--1.5~T. Our estimates therefore apply to the lower part of the experimental field range. Treating the defect as a local perturbation of the exchange or of the anisotropy, as done in Ref.~\cite{Hanneken2016}, our effective model shows that the curvature of the landscape increases by a factor 1.7--3.3 when the skyrmion shrinks from 3.9 to 1.6~nm, i.e.\ when the field is raised from 1 to 3~T. Quantum effects are therefore enhanced at the highest fields compatible with isolated skyrmions, around 3~T, where the level spacings and crossover temperatures of Table~\ref{tab:maps} could increase by up to about a factor of two, and where the magnon gap of about 1~meV freezes magnon-mediated damping.

Within STM, the tip itself offers also an elegant possibility. The local exchange and stray fields of a spin-polarized tip, as well as the tip height, tilt the energy landscape in a controlled fashion, while dragging a cluster~\cite{Hanneken2016} displaces the pinning potential itself. The escape of the skyrmion from its pinning site, or its switching between two neighboring minima, can then be monitored as telegraph noise in the tunneling current with the tip parked at low bias; noise spectroscopy has already proven to be a sensitive probe of skyrmion motion~\cite{Sato2019}. Rates of 1--100~Hz, as found in Sec.~\ref{sec:depinning}, are ideal for such time-resolved measurements. Above $T^*$, the switching rate follows the Arrhenius law of thermal activation, as expected from stochastic Thiele equations~\cite{Troncoso2014} and observed for skyrmion creep~\cite{Luo2020}; below $T^*$, it becomes temperature independent, which is the most robust signature of quantum motion.

Electron spin resonance STM (ESR-STM)~\cite{Baumann2015} offers a more direct, spectroscopic route, as also proposed for the readout of skyrmion helicity states~\cite{PsaroudakiPanagopoulos2022}. The gyrotropic level spacings of the pinned skyrmion, 5--20~GHz for the adatoms and dimers of Table~\ref{tab:maps}, fall precisely within the frequency window of ESR-STM. A radio-frequency voltage applied to the junction modulates the tip-induced potential and thus exerts an oscillating force on the skyrmion, driving transitions between neighboring orbital levels. Indeed, a uniform oscillating force $\mathbf F_{\mathrm{rf}}(t)$ enters as $-\mathbf F_{\mathrm{rf}}\cdot\hat{\mathbf R}$, and since $\hat X=\ell(\hat a+\hat a^\dagger)/\sqrt2$ in an isotropic well, it couples only levels with $\Delta n=\pm1$, with a Rabi energy $\hbar\Omega=F_{\mathrm{rf}}\ell\sqrt{n+1}/\sqrt2$. A modulation of the tip-induced potential by 0.1~meV over 5~\AA\ yields $\hbar\Omega\approx9~\mu$eV, comparable to the linewidth $\hbar\omega/Q_{\mathrm f}$, such that the transitions can be driven efficiently. The resulting change of the time-averaged position of the skyrmion is detected through the spin-polarized tunneling current, similarly to the detection of the spin resonance of single atoms. Such a measurement would provide three pieces of information. First, the resonance frequency gives directly the curvature of the energy landscape, $\hbar\omega=\ell^2\sqrt{k_xk_y}$, allowing a defect-by-defect test of the \textit{ab initio} landscapes. Second, the linewidth measures the damping; with $Q_{\mathrm f}\approx5$--50, the resonance can be resolved in frequency sweeps. Third, the temperature dependence reveals the quantum regime. We note that for a harmonic landscape the quantum and classical gyrotropic frequencies coincide, as for any harmonic oscillator. The quantum nature manifests itself through the anharmonicity of the landscape: thermal occupation of higher orbitals shifts and broadens the resonance, and below $T^*$ this shift and broadening saturate at their zero-point values. Two challenges should be kept in mind. ESR-STM has so far been applied mostly to atoms on thin insulating layers, and the coupling between the radio-frequency field and the skyrmion position on a metallic substrate remains to be quantified. Yet, the classical gyrotropic resonance of magnetic vortices is routinely excited by microwaves, and its skyrmionic analogue at the atomic scale would by itself constitute an important milestone.

A skyrmion positioned on top of a repulsive defect would, in addition, survive for several orbital periods, with a lifetime settled by the damping rather than by the shape of the hill. All these experiments require reading the skyrmion position with sub-\AA ngstr\"om resolution without perturbing it. The spin-mixing magnetoresistance is helpful in this respect: at the vicinity of implanted defects it varies on the \aa ngstr\"om scale, and early $4d$ impurities enhance the tunneling signal by up to about 350\%~\cite{LimaFernandes2020NC}, such that the defect pinning the skyrmion can also act as its position sensor. The choice of the bias voltage is, however, critical. At $+0.7$~V, where the non-collinear magnetoresistance is large, the injected electrons are energetic enough to write and delete skyrmions~\cite{Hanneken2016}, whereas at a bias of a few mV even currents of tens of nA leave the skyrmion intact~\cite{Hanneken2016}. Low-bias detection with a spin-polarized tip, exploiting the tunneling magnetoresistance, is therefore preferable, and the tip--skyrmion interaction itself has to be accounted for as part of the energy landscape. Since the tip potential can reach the meV scale, it should be kept smooth on the scale of $\ell$ and weak compared to the defect landscape, by working at large tip--sample distances; a controlled tip potential is, in turn, precisely what enables the tilt of the landscape discussed above.

\section{Discussion and conclusions}
\label{sec:outlook}

Our investigations bear some limitations, which define natural next steps. The energy landscapes were extracted from published maps, interpolated from a finite grid of skyrmion positions, which limits the accuracy of the curvatures; we quantified this uncertainty, which remains moderate except for the top of the Cr hill and the Cr saddle. Dense landscapes computed directly would sharpen our predictions, while a joint quantization of the position and of the internal modes would extend the theory to the strongly interacting embedded impurities found in ab-initio  calculations. Finally, a first-principles evaluation of the damping of the skyrmion-defect complex would refine our estimates of the quality factor.

Our findings also bear on the interpretation of geodesic nudged elastic band (GNEB) simulations~\cite{Bessarab2015}, which are commonly utilized to evaluate energy barriers and lifetimes of skyrmions. We note that the barriers and wells discussed here are skyrmion-defect interaction energies, not the barriers for nucleation or collapse of the skyrmion. GNEB calculations remain, however, the natural input of our approach: the barrier height and the curvatures at the minimum and at the saddle point determine both the classical escape rate and the quantum exponent $B$ of Eq.~\eqref{eq:B}. Below the crossover temperature $T^*$, rates extrapolated from harmonic transition-state theory underestimate the actual escape rate, since the skyrmion tunnels in phase space instead of being thermally activated over the saddle point. Combining GNEB with the quantum treatment, as done for skyrmion collapse~\cite{Vlasov2020}, offers a natural route to quantitative lifetimes at millikelvin temperatures.

Our findings open appealing perspectives beyond PdFe/Ir(111). Magnetic nanostructures are not passive: the Cr dimer switches its spin chirality when approached by a skyrmion, and frustrated Cr trimers modify their non-collinear state~\cite{GedeArjana2020,Lounis2007}. Moments following the skyrmion adiabatically contribute a Berry phase, which locally renormalizes $G$ and thus $\ell$. Where adiabaticity breaks down, at the vicinity of such switches, the skyrmion becomes coupled to a discrete quantum spin, which could pave the way toward entangling the position of a skyrmion with a localized moment. In antiferromagnetic, synthetic antiferromagnetic or altermagnetic hosts, the gyrotropic force vanishes altogether. The skyrmion becomes then a regular massive particle, able to tunnel coherently between the two minima of a dimer. In frustrated magnets, the helicity of the skyrmion offers yet another quantum degree of freedom, which was proposed as the basis of skyrmion qubits~\cite{Psaroudaki2021,PsaroudakiPanagopoulos2022,PsaroudakiAPL2023}. Interestingly, defect engineering is required there to shape the double-well potential of the helicity~\cite{Psaroudaki2021}. We envision designed defects as a natural route toward skyrmionic two-level systems.

To summarize, the gyrotropic force transforms a ferromagnetic skyrmion into a particle living in the lowest Landau level, with a magnetic length of 0.6~\AA, and its defect-induced energy landscape into its Hamiltonian. For PdFe/Ir(111), we predict quantized gyrotropic orbits in both pinning wells and repulsive hills, with spacings of 20--85~$\mu$eV, saddle points acting as quantum beam splitters and quantum depinning at rates of 1--100~Hz below the critical current. Embedded impurities, which interact much more strongly with the skyrmion, define instead a strong-coupling regime in which position and internal modes have to be treated jointly. For the damping typical of ultrathin films, the quantized levels remain well resolved. The thermal-to-quantum crossover lies between 40 and 160~mK. The defects maximizing these effects can be crafted atom by atom, and tip-induced escape statistics as well as ESR-STM, at millikelvin temperatures and fields around 3~T, offer realistic routes to reveal them. Our work promotes atomic defects from inhibitors of skyrmion motion to building blocks of quantum skyrmionics.

\begin{acknowledgments}
The author acknowledge discussions with Imara Lima Fernandes. This work was supported by the Deutsche Forschungsgemeinschaft (DFG) through Project No.~328545488 -- CRC/TRR 227, Project No.~B12, and by the German Excellence Strategy -- EXC 3112/1 -- 533767171 (Center for Chiral Electronics). The author discloses the use of AI tools to assist with drafting and refining portions of the manuscript based on author-provided scientific ideas and context that were critically reviewed, verified, and revised.
\end{acknowledgments}


\appendix

\section{LLL spectrum of a radial landscape}
\label{app:wigner}

With $[\hat X,\hat Y]=i\ell^2$, define $\hat a=(\hat X+i\hat Y)/(\sqrt2\,\ell)$. Then $\hat X^2+\hat Y^2=\ell^2(2\hat a^\dagger\hat a+1)$, and the eigenstates of a rotationally symmetric Weyl-ordered Hamiltonian are the number states $|m\rangle$. Their Wigner function in the $(X,Y)$ phase plane is $W_m$ of Eq.~\eqref{eq:Wigner}, and $E_m=\int d^2r\,W_mU$ exactly. Numerically, $e^{-x/2}L_m(x)$ with $x=2r^2/\ell^2$ is generated by the stable three-term Laguerre recurrence, and the radial integral is evaluated on a grid of $4\times10^4$ points up to 16~\AA\ for $m\le 260$. For $U=kr^2/2$ the implementation reproduces $E_m=k\ell^2(m+\frac12)$ to machine precision.

\section{Escape exponent}
\label{app:escape}

In the LLL, the two in-plane coordinates are conjugate with $\hbar_{\mathrm{eff}}=\ell^2$. Let $s$ be the coordinate along the drive and $t$ the transverse one. The classically forbidden region at the energy $E$ of the metastable state is spanned by imaginary $t$, and with the local transverse curvature $k_\perp(s)$ the WKB exponent is
\begin{equation}
B = \frac{2}{\ell^2}\int_{s_1}^{s_2} \sqrt{\frac{2\,[V(s) - E]}{k_\perp(s)}}\; ds, \qquad V(s) = U_{\mathrm{eff}}(s, 0).
\label{eq:B1}
\end{equation}
For a radial landscape, $k_\perp(s)=U'(|s|)/|s|$. For the cubic form of Sec.~\ref{sec:depinning} ($s=x$, $k_\perp=k_y$), measure $s$ from the well: $V-V_{\min}=\lambda(\sqrt\delta\,s^2-s^3/3)$ on $0<s<3\sqrt\delta$. At $E=V_{\min}$, Eq.~\eqref{eq:B1} gives
\begin{equation}
B=\frac{2}{\ell^2}\sqrt{\frac{2\lambda}{k_y}}\;9\,\delta^{5/4}\,\mathrm{B}(2,\tfrac32)=\frac{24}{5}\sqrt{\frac{2\lambda}{k_y}}\,\frac{\delta^{5/4}}{\ell^2},
\end{equation}
with the Beta function $\mathrm{B}(2,\frac32)=4/15$. Using $\Delta U_b=\frac43\lambda\delta^{3/2}$ and $\hbar\omega_w=\ell^2\sqrt{2\lambda\sqrt\delta\,k_y}$, this equals $(36/5)\Delta U_b/\hbar\omega_w$. For the inatom landscapes Eq.~\eqref{eq:B1} is integrated numerically; it converges to the cubic result for $1-u/u_c\lesssim0.05$.


\end{document}